\documentstyle[12pt,a4wide]{article}
\input epsf

\newcommand{\news}{\setcounter{equation}{0} \ \indent}
\newcommand{\be}{\begin{equation}}
\newcommand{\ee}{\end{equation}}
\newcommand{\bea}{\begin{eqnarray}}
\newcommand{\eea}{\end{eqnarray}}
\newcommand{\bean}{\begin{eqnarray*}}
\newcommand{\eean}{\end{eqnarray*}}
\font\upright=cmu10 scaled\magstep1
\font\sans=cmss12
\newcommand{\ssf}{\sans}
\newcommand{\stroke}{\vrule height8pt width0.4pt depth-0.1pt}
\newcommand{\Z}{\hbox{\upright\rlap{\ssf Z}\kern 2.7pt {\ssf Z}}}

\newcommand{\C}{{\rlap{\rlap{C}\kern 3.8pt\stroke}\phantom{C}}}
\newcommand{\R}{\hbox{\upright\rlap{I}\kern 1.7pt R}}
\newcommand{\CP}{\C{\upright\rlap{I}\kern 1.7pt P}}

\newcommand{\half}{\frac{1}{2}}

\newcommand{\identity}{{\upright\rlap{1}\kern 2.0pt 1}}

\begin{document}
\pagestyle{plain}
\title{
\begin{flushright}
{\normalsize UKC/IMS/96-20}\\ 
{\normalsize To appear in Physics Letters B} \\
\end{flushright}
\vskip 20pt
{\bf Monopole Zeros} \vskip 20pt}
\author{Paul M. Sutcliffe
\thanks{This work was supported in part by the 
Nuffield  Foundation}\\[20pt]
{\sl  Institute of Mathematics} \\[5pt]
{\sl University of Kent at Canterbury} \\[5pt]
{\sl  Canterbury CT2 7NZ, England} \\[20pt]
{\sl email\  P.M.Sutcliffe@ukc.ac.uk }\\[10pt]}

\date{January 1996}

\maketitle

\begin{abstract}
Recently the existence of certain SU(2) BPS monopoles
with the symmetries of the Platonic solids has been proved.
Numerical results in an earlier paper suggest
that one of these new monopoles, the tetrahedral 3-monopole,
has a remarkable new property, in that the number of
zeros of the Higgs field is greater than the topological
charge (number of monopoles). As a consequence, zeros
of the Higgs field exist (called anti-zeros)
around which the local winding
number has opposite sign to that of the total winding.
In this letter we investigate the presence of anti-zeros
for the other Platonic monopoles. Other aspects of 
anti-zeros are also discussed.

\end{abstract}
\newpage

\section{Introduction}
\news
SU(2) BPS monopoles are topological soliton solutions
of a Yang-Mills-Higgs gauge theory in three space
dimensions. They are Bogomolny solitons, in that
they attain a topological lower bound on the total
energy, and so can be obtained as solutions of a 
first order equation (the Bogomolny equation)
 rather than the more general second order field equations.
In this letter we shall mainly be concerned with 
static monopoles. 
The ingredients of the static theory are the Higgs field
$\Phi$, and the gauge potential
$A_i, \ i=1,2,3$, both of which are {\sl su(2)}-valued.
The static theory can be defined by its energy density
\be
{\cal E}=-\half\mbox{tr}(D_i\Phi)(D_i\Phi)-\frac{1}{4}
\mbox{tr}(F_{ij}F_{ij})
\label{energy}
\ee
where $D_i=\frac{\partial}
{\partial x_i}+[A_i,$ \ is the covariant derivative
and $F_{jk}$ the gauge field. Integrating the energy
density over all \R$^3$ gives the energy $E$ 
of any configuration. 

The boundary condition
\be
\|\Phi\|\rightarrow 1 \hskip 10pt\mbox{as}
\hskip 10pt r\rightarrow\infty
\label{bc}
\ee
where $r=\vert\mbox{\boldmath $x$}\vert$, 
$\|\Phi\|^2=-\half\mbox{tr}\Phi^2$, is imposed
and may be thought of as a residual
finite energy condition derived from a vanished
Higgs potential.

The topological aspect arises because 
the Higgs field at infinity induces a map between spheres:
\be\Phi:S^2(\infty)\rightarrow S^2(1)\ee
where $S^2(\infty)$ is the two-sphere at spatial infinity and
$S^2(1)$ is the two-sphere of vacuum configurations given by
$\{\Phi\in su(2) : \|\Phi\|=1\}.$
The degree of this map is an integer $k$, the 
winding number, which
(in suitable units) is the total magnetic charge of the
monopole. 
We shall refer to a monopole
with magnetic charge $k$ as a $k$-monopole. 

The Bogomolny bound
$$E\ge 8\pi \vert k\vert$$ gives a lower
bound on the total energy of a configuration in
terms of its winding number $k$.
For each $k$ this bound can be attained, with
the relevant configuration being a solution
of the first order Bogomolny equation
\be
D_i\Phi=\pm\half\epsilon_{ijk}F_{jk}
\label{bog}
\ee
where the lower sign corresponds to positive
$k$, with solutions called monopoles, and
the upper sign corresponds to $k$ being 
negative, where solutions are called anti-monopoles.
We now choose to consider monopoles ie. $k>0$, and
so fix the sign in (\ref{bog}) to be the lower one.

By topological considerations 
the total number of zeros 
of the Higgs field  {\sl counted with
 multiplicity}
is $k$ for a $k$-monopole.
These zeros need not be distinct, for example, 
axially symmetric monopoles exist for all $k\ge 2$
\cite{W,P}
for which there is a single zero that has multiplicity $k$.
When the zeros are distinct, and well separated, the
$k$-monopole solution has a natural interpretation
as $k$ well-separated unit charge monopoles, each
one centered at a zero of the Higgs field. 
Such solutions exist because physically there
are no static forces between equally charged monopoles
\cite{Ma}. 
The moduli space of charge $k$ monopoles
is $4k$ dimensional \cite{CG}, and this is again consistent
with the well-separated picture, where each
individual charge one monopole has three position
degrees of freedom and one internal phase.

So the general picture appears to be that the Higgs
field of a charge $k$ monopole has $k$ simple zeros,
which may be thought of as the location of each
monopole, and these zeros can coalesce as the monopoles
merge. Although this picture has never been 
rigorously proved, it is widely accepted as true.
Indeed there are at least three compelling reasons 
for believing the above. Firstly, all the known explicit
monopole solutions do indeed have a Higgs field of
this form.
Secondly, in the analogous two dimensional case of
abelian Higgs vortices at critical coupling, it has been
proved that the total number of zeros of the Higgs
field is bounded by the number of vortices \cite{JT}.
Furthermore, in other models with topological solitons,
such as the O(3) $\sigma$-model in the plane, the general
static
$k$-soliton solution can be given explicitly  and
a suitable field shown to have the above structure of zeros
\cite{BP}.
Finally, if the total number of Higgs zeros
(ignoring multiplicities) were greater than $k$ then
this would imply that zeros with {\sl negative
multiplicity} must exist, for the summed multiplicities
to equal $k$. 
From now on we shall refer to a zero with
a negative multiplicity as an anti-zero.
An example of a configuration with
an anti-zero is of course a single anti-monopole,
which has $k=-1$. So it would seem highly unlikely
that a monopole configuration with $k>0$ could contain
anti-zeros which were well-separated from other zeros,
since we could interpret such a configuration as
composed of monopoles and anti-monopoles. Since there
are attractive forces between monopoles of opposite
charge such a configuration could not saturate the
Bogomolny energy bound.

Despite this wealth of circumstantial evidence,
it was argued in a recent paper \cite{HSc},
 with the aid numerical results,
that a positive charge monopole solution exists which
contains anti-zeros. 
In this letter we briefly review this result for
the tetrahedral 3-monopole and then investigate the
 presence of anti-zeros for the remaining Platonic monopoles.
Other aspects of 
anti-zeros will also be discussed, such as a signal for
their occurrence and their relevance to skyrmions.

\section{Zeros of Platonic Monopoles}
\news
Recently it has been shown that monopoles exist which
have the symmetries of the Platonic solids \cite{HMM,HSb}.
The actual monopole fields $\Phi,A_i$, were not 
calculated explicitly but rather a twistor approach
was taken in which monopole solutions can be shown to be
equivalent to certain algebraic objects, called
spectral curves \cite{Ha}. The spectral curves were explicitly
found, from which the existence and symmetries of
the monopoles follows. Using a numerical implementation
of the twistor transform \cite{HSa}, the Higgs field
and energy density of these monopoles can be computed.
The results were displayed graphically in the form
of a three-dimensional plot of a surface of constant 
energy density. For each of the four newly discovered
monopoles it was found that a surface of constant
energy density resembled a Platonic solid. The results
are summarized in Table 1, where we give the monopole
charge $k$ and the Platonic solid it resembles.
In each case the energy density takes its maximum
values on the vertices of the relevant Platonic solid.

\begin{center}
\begin{tabular}{|c|c|} \hline
Charge k & Platonic solid \\
\hline
3 & Tetrahedron \\
4 & Cube \\
5 & Octahedron \\
7 & Dodecahedron \\
\hline
\end{tabular}
\end{center}
\begin{center}
Table 1. {\sl Charges of  Platonic Monopoles }
\end{center}\ 
In these papers the Higgs field and its zeros
were not studied, since if there are $k$ zeros
then in each case the symmetry group acting
implies that all $k$ zeros must be at the origin.
For example, for the tetrahedral monopole $k=3$ and if
there are only three zeros then in order to arrange
three points with tetrahedral symmetry all three points must
be at the origin.

However, using the moduli space approximation \cite{M,S}
the dynamics of $k$ slowly moving monopoles can be approximated
by geodesic motion on the monopole moduli space 
${\cal M}_k$. 
In \cite{HSc} we presented a totally geodesic
one dimensional submanifold of the 3-monopole moduli
space, which contains on it the tetrahedral 3-monopole.
This geodesic may therefore be interpreted in terms
of the scattering of three monopoles which instantaneously
form the tetrahedral 3-monopole. These results 
make it appear very unnatural (see \cite{HSc} for
more details) that there are three zeros at the origin and
so we examined the Higgs field in more detail.
Writing the Higgs field in terms of Pauli
matrices as
\be
\Phi=i\sigma_1\varphi_1+i\sigma_2\varphi_2
+i\sigma_3\varphi_3
\ee
we plotted the components $\varphi_1,\varphi_2,\varphi_3$
along the line $x_1=x_2=x_3=L$, 
which goes through a vertex (at a negative value of $L$)
and the center of a face (at a positive value of $L$)
of the tetrahedron associated with the tetrahedral 
monopole.
Fig 1. shows the results, and it is clear that along
this line there are two points at which all the
component of the Higgs field vanish. One point is
the origin ($L=0$) and the second occurs at a negative
value of $L$, which indicates it is associated with
a vertex of the tetrahedron, rather than a face.
There are another three similar lines, going through
the remaining vertices of the tetrahedron, and these
were the only other lines along which Higgs zeros were
found. So the result is that there are five Higgs zeros,
one associated with each vertex of the tetrahedron and
one at the origin. Since the monopole charge is three,
then the zero at the origin must be an anti-zero.
This can be checked numerically (see \cite{HSc} for details
of the scheme) by computing the winding number 
$Q(r)$, of the unit 3-vector
\be
\psi=(\varphi_1,\varphi_2,\varphi_3)
\frac{1}{\sqrt{\varphi_1^2+\varphi_2^2+\varphi_3^2}}
\ee
corresponding to the normalized Higgs field
on a two-sphere of radius $r$, centred
at the origin. 
Note that by definition $Q(R)=k$, if $R$ is sufficiently
large, so that all zeros of the Higgs field are
contained within the ball of radius $R$
centred at the origin. Such a calculation gives
$Q(0.2)=-1$ and $Q(1.0)=+3$, confirming that there
is indeed an anti-zero at the origin.

Having briefly reviewed the results for the tetrahedral
monopole we now go on to investigate the Higgs zeros
of the other Platonic monopoles.

We begin with the cubic 4-monopole. First of all we
explicitly prove that the Higgs field of the cubic monopole 
is zero at the origin. It is useful to give the details
of this calculation, since it demonstrates the kind
of work required to prove the results
which the numerical evidence suggests.

Recall the ADHMN construction
\cite{N,Hb} which is an equivalence between $k$-monopoles 
and Nahm data
$(T_1,T_2,T_3)$, which are three $k\times k$
 matrices which depend
on a real parameter $s\in[0,2]$ and satisfy the following;\\

\newcounter{con}
\setcounter{con}{1}
(\roman{con})  Nahm's equation
\be
\frac{dT_i}{ds}=\half\epsilon_{ijk}[T_j,T_k] \nonumber
\ee\\

\addtocounter{con}{1}
(\roman{con}) $T_i(s)$ is regular for $s\in(0,2)$ and has simple
poles at $s=0$ and $s=2$,\\

\addtocounter{con}{1}
(\roman{con}) the matrix residues of $(T_1,T_2,T_3)$ at each
pole form the irreducible $k$-dimensional representation of SU(2),\\

\addtocounter{con}{1}
(\roman{con}) $T_i(s)=-T_i^\dagger(s)$,\\

\addtocounter{con}{1}
(\roman{con}) $T_i(s)=T_i^t(2-s)$.\\

Finding the Nahm data effectively solves the nonlinear part 
of the monopole construction
and is enough to prove existence of the monopole and
compute its spectral curve.
In fact this is how
the spectral curves of the Platonic monopoles were
calculated.
However in order to calculate the Higgs field
the linear part of the ADHMN construction must also be
 implemented. Given
Nahm data $(T_1,T_2,T_3)$ for a $k$-monopole we must solve the 
ordinary differential equation  
\be
({\identity}_{2k}\frac{d}{ds}+{\identity}_k\otimes x_j\sigma_j
+iT_j\otimes\sigma_j){\bf v}=0
\label{lin}
\ee
for the complex $2k$-vector ${\bf v}(s)$,
 where $\identity_k$ denotes
the $k\times k$ identity matrix, $\sigma_j$ are the
 Pauli matrices and
${\bf x}=(x_1,x_2,x_3)$ is the point in space at
 which the Higgs
field is to be calculated. Introducing the inner product
\be
\langle{\bf v}_1,{\bf v}_2\rangle =\int_0^2 {\bf v}_1^\dagger{\bf v}_2\ ds
\label{ip}
\ee
then the solutions of (\ref{lin}) which we require are
 those which are
normalizable with respect to (\ref{ip}). It can be shown
 that the
space of normalizable solutions to (\ref{lin}) has 
(complex) dimension
2. If $\widehat {\bf v}_1,\widehat {\bf v}_2$ 
is an orthonormal basis
for this space then the Higgs field $\Phi$ is given by
\be
\Phi=i\left[ \begin{array}{cc}
\langle(s-1)\widehat {\bf v}_1,\widehat {\bf v}_1\rangle &
\langle(s-1)\widehat {\bf v}_1,\widehat {\bf v}_2\rangle \\
\langle(s-1)\widehat {\bf v}_2,\widehat {\bf v}_1\rangle &
\langle(s-1)\widehat {\bf v}_2,\widehat {\bf v}_2\rangle 
\end{array}
\right].
\label{higgs}
\ee
For the cubic monopole $k=4$ and the Nahm data is
explicitly known \cite{HMM}. Writing 
${\bf v}=(v_1,v_2,v_3,v_4,v_5,v_6,v_7,v_8)^t$ then
(\ref{lin}) becomes the set of equations
\bea
& &\dot v_1+x_3v_1+(x_1+ix_2)v_2
+(4y+3x)v_1+20yv_8=0
\nonumber\\
& &\dot v_2-x_3v_2+(x_1-ix_2)v_1
-(4y+3x)v_2+2\sqrt{3}(-2y+x)v_3=0
\nonumber\\
& &\dot v_3+x_3v_3+(x_1+ix_2)v_4
+2\sqrt{3}(-2y+x)v_2+(-12y+x)v_3=0
\nonumber\\
& &\dot v_4-x_3v_4+(x_1-ix_2)v_3
+(12y-x)v_4+4(3y+x)v_5=0
\nonumber\\
& &\dot v_5+x_3v_5+(x_1+ix_2)v_6
+4(3y+x)v_4+(12y-x)v_5=0
\nonumber\\
& &\dot v_6-x_3v_6+(x_1-ix_2)v_5
+(-12y+x)v_6+2\sqrt{3}(-2y+x)v_7=0
\nonumber\\
& &\dot v_7+x_3v_7+(x_1+ix_2)v_8
+2\sqrt{3}(-2y+x)v_6-(4y+3x)v_7=0
\nonumber\\
& &\dot v_8-x_3v_8+(x_1-ix_2)v_7
+20yv_1+(4y+3x)v_8=0.
\eea
where dot denotes differentiation with
respect to $s$
\begin{eqnarray} x(s)&=&\frac{\kappa}{5}\left(-2\sqrt{\wp(\kappa
    s)}+\frac{1}{4}\frac{\wp^\prime(\kappa s)}{\wp(\kappa
    s)}\right)\\
y(s)&=&\frac{\kappa}{20}\left(\sqrt{\wp(\kappa
    s)}+\frac{1}{2}
\frac{\wp^\prime(\kappa s)}{\wp(\kappa s)}\right)
\end{eqnarray}
with $\kappa$ a known constant 
and $\wp$ the elliptic function satisfying
\be \wp^{\prime 2}=4\wp^3-4\wp.\ee
Here prime denotes differentiation with respect
to the argument. The constant 
$\kappa=\Gamma(1/4)^2/\sqrt{8\pi}$ is such that
the real period of the elliptic function is $2\kappa$.

We wish to calculate the Higgs field at the origin
so we set  $x_1=x_2=x_3=0$. Then 
the first and last equations 
decouple from the rest, so we may look for
a solution with $v_2=v_3=v_4=v_5=v_6=v_7=0$ to give
\bea
& &\dot v_1
+(4y+3x)v_1+20yv_8=0
\nonumber\\
& &\dot v_8
+(4y+3x)v_8+20yv_1=0.
\eea
The symmetry of this system allows the reduction $v_8=v_1$
which brings us to the single equation
\be
\dot v_1+(24y+3x)v_1=0.
\ee
Now 
\be
24y+3x=\frac{3\kappa}{4}\frac{\wp^\prime}{\wp}=\frac{3}{4}
\frac{\dot\wp}{\wp}
\ee
so the equation is
\be
\dot v_1+\frac{3}{4}\frac{\dot\wp}{\wp} v_1=0
\ee
with solution
\be
v_1=A\wp^{-3/4}
\ee
where $A$ is a constant.
The properties of the elliptic function
$\wp$ are such that $v_1$ is finite
for $s\in[0,2]$. Hence we have our first
unit norm solution
\be
\widehat{\bf v}_1=B^{-1}\wp^{-3/4}(1,0,0,0,0,0,0,1)^t
\ee
where $B$ is the constant
\be 
B^2=2\int_0^2 \wp^{-3/2}\ ds.
\ee
In a similar way the fourth and fifth equations
in (\ref{lin}) decouple to give
\be
\widehat{\bf v}_2=B^{-1}\wp^{-3/4}(0,0,0,1,1,0,0,0)^t.
\ee 
Substituting these solutions into (\ref{higgs})
we have that
\be
\Phi=i2B^{-2} \identity_2 \int_0^2
 (s-1)\wp^{-3/2}\ ds.
\label{phizero}
\ee
However, $\wp$ is symmetric on the real line about its half
period, so the integrand in (\ref{phizero}) is 
antisymmetric about $s=1$, and hence $\Phi=0$.
So finally, we have proved that the Higgs field of
the cubic monopole has a zero at the origin.
To find the Higgs field at points other than the
origin requires a similar calculation, but it is
more involved, since for a general point the set of equations
will no longer decouple in a simple way.
This is why it is a difficult task to prove that the
tetrahedral 3-monopole has five zeros, and instead we
rely on numerical results.

Essentially the numerical scheme \cite{HSa} solves
the linear differential system (\ref{lin}), extracts
an orthonormal basis for the normalizable solutions
and performs the required integrations to obtain
the Higgs field.

Returning to a numerical investigation of the
cubic monopole we plot, in Fig 2. (the solid curve),
the norm squared of the Higgs field $\|\Phi\|^2$
along the line $x_1=x_2=x_3=L$. This line goes through
two vertices of the cube associated with the cubic
monopole. It is not easy to see exactly where this
function is zero, so we also plot in Fig 3. the
component $\phi_2$ (solid curve). From this plot
it seems relatively clear that the cubic monopole
has a zero of the Higgs field only at the origin,
and not at points associated with the vertices of
a cube. Calculation of the Higgs field along other
lines, for example through the centre of a face,
did not reveal any other zeros. So it would seem
that the cubic monopole is not like the tetrahedral
monopole, and does not possess anti-zeros.
Supporting evidence comes from a calculation of
the winding number around the origin, which
gives $Q(0.1)=+4$, in agreement with the cubic
monopole having just four zeros, which are all
located at the origin.

Having demonstrated that the tetrahedral monopole appears 
to have an anti-zero, it may now seem surprising that the 
cubic monopole has no anti-zeros. However, there are 
several reasons why we might expect the cubic
monopole to be unlike the tetrahedral monopole. The first
is obvious, in that if the cubic monopole had a zero
at each vertex then, since it has charge four, there
would have to be four anti-zeros at the origin ie. an
anti-zero with local winding -4. This requirement of
multiple anti-zeros is not impossible, but it does
seem a little contrived.

A second reason comes from the study of monopole
scattering, which can be addressed using the moduli
space approximation, as mentioned earlier.
There is a totally geodesic one dimensional submanifold
of ${\cal M}_3$ which contains the tetrahedral monopole
\cite{HSc}. The Nahm data associated with this submanifold
involves the family of elliptic curves
\be
y^2=4x^3-3(a^2-4)^{2/3}x-4
\label{curvea}
\ee
where $a\in\R$ is a parameter, such that $a=\pm 2$
gives the tetrahedral monopole. Now there are points
on the geodesic which correspond to three well-separated
unit charge monopoles, so we know that at such points
the corresponding monopole configuration can have no
anti-zeros. Since the tetrahedral monopole has anti-zeros
there must be special  \lq splitting points\rq\ along
the geodesic at which anti-zeros appear/disappear.
The discriminant of the elliptic curve (\ref{curvea})
is 
\be
\Delta=27a^2(a^2-8)
\ee
which vanishes at the three points $a=0,\pm\sqrt{8}$.
The numerical results are consistent with the
conjecture that the splitting points occur at these
three parameter values at which the elliptic curve 
is singular. 

There is a geodesic in ${\cal M}_4$ which
contains the cubic monopole \cite{HSa,Sa} and is associated
with the family of elliptic curves
\be
y^2=4x^3-4x+12a^2
\label{curveb}
\ee
where $a\in(-3^{5/4}\sqrt{2},+3^{5/4}\sqrt{2})$.
The discriminant of this elliptic curve
is 
\be
\Delta=16(4-3^5a^4)
\ee
which never vanishes for $a$ in its allowed range.
Hence, if the above conjecture is correct, then there
are no splitting points along this geodesic. Since the
geodesic contains points which correspond to four
well-separated unit charge monopoles, then this implies
that the cubic monopole has no anti-zeros.
In Fig 2. and Fig 3. we plot $\|\Phi\|^2$ and $\varphi_2$
(dashed lines) along the line $x_1=x_2=x_3=L$ for
two monopole configurations corresponding to
two other points on the geodesic in ${\cal M}_4$.
From these (and similar) plots it can be seen
that in each case there is only one zero along
this line, which can be tracked as it moves in from
infinity, through the origin and back out to infinity
in the opposite direction along the line.
There is no signature of any splitting of zeros
taking place.

A final indication that the tetrahedral and cubic
monopoles have a different structure in their zeros
comes from an analogy with another kind of topological
soliton, the skyrmion. Numerical evidence suggests
\cite{BTC} that the minimal energy 3-skyrmion and the minimal
energy 4-skyrmion resemble a tetrahedron and a cube
respectively, so there is some similarity between
monopoles and skyrmions. Using instanton generated
skyrmions \cite{LM} these kinds of configurations were
investigated in more detail, and it was found that
the tetrahedral skyrmion has regions in which the
 baryon density is negative, but no such regions were
 found for the cubic skyrmion. For skyrmions the 
baryon density is the quantity which when integrated
over all \R$^3$ gives the number of skyrmions ie. it
is the topological charge density. For an anti-skyrmion
the baryon density is negative, so for a skyrmion
configuration with positive topological charge to
have a region in which the baryon density is negative
is analogous in the monopole context to a region 
containing an anti-zero. Hence, if the tetrahedral
monopole contains anti-zeros, but the cubic monopole
does not then this is yet again another parallel
between monopoles and skyrmions.

Having looked for anti-zeros in the tetrahedral
and cubic monopoles and finding apparently different
answers for each, it is by no means clear what
the situation will be for the other two Platonic monopoles.
Note that for the octahedron, since the charge is five, 
a zero at each vertex would imply that only a single
anti-zero is required at the origin. Thus in this respect
the octahedral monopole is like the tetrahedral 
monopole rather than the cubic monopole, and is
a candidate for anti-zeros. Fig 4. shows a plot of
$\|\Phi\|^2$ for the octahedral monopole, along the
line $x_1=x_2=0, \ x_3=L$, which passes through 
two vertices of the associated octahedron.
This clearly suggests that there are three zeros
along this line. There are two other similar lines,
so we find that the octahedral monopole has a zero on each
of the six vertices of the octahedron and an anti-zero
at the origin. This conclusion is supported by a
winding number calculation which gives $Q(3.0)=+5$ and
$Q(0.1)=-1$.

Numerical results for the dodecahedral 7-monopole
are not as conclusive as for the other three Platonic
monopoles, but seem to suggest that it is like the
cubic 4-monopole in not possessing anti-zeros.
This would seem the most acceptable result, since
if the dodecahedral monopole had anti-zeros in the
same manner as the tetrahedral and octahedral monopole
then this would require multiple anti-zeros 
(in fact thirteen) at the origin. 

It would clearly be desirable to test the conjecture,
relating splitting points to singular elliptic
curves, with other examples. In particular it
would be instructive if Nahm data could be found
which corresponds to a geodesic in ${\cal M}_5$
that includes the octahedral monopole. The
conjecture implies that the associated family
of elliptic curves should contain singular curves.
Two appropriate one-dimensional totally geodesic
submanifolds of  ${\cal M}_5$ are known \cite{HSb,HSc},
but unfortunately the computation of the associated
Nahm data appears not to be a tractable problem.
However, a more suitable candidate does appear to
exist and is obtained by imposing tetrahedral
symmetry on five monopoles. This should be 
investigated as it could provide a simple
counter example to prove the conjecture false,
if it could be shown that such a geodesic exists
and its associated family of elliptic curves
contained no singular curves.

\section{Conclusion}
\news
In this letter we point out that it appears that
monopole solutions exist which saturate the
Bogomolny energy bound and yet which have more zeros
of the Higgs field than number of monopoles.
We refer to such spurious zeros as anti-zeros,
since they have a local winding which has
opposite sign to the total charge of the monopole.
 Whether such monopole configurations
could be interpreted as BPS monopole anti-monopole 
states is not yet known, since such an interpretation
would require a local definition of magnetic charge
density (because the zeros and anti-zeros are close
together). At present no useful definition exists,
since the standard definition relies upon a 
consideration of the asymptotic field far from the
monopole where the non-abelian symmetry is broken
to an abelian symmetry which can be identified
with electromagnetism.

Some discussion on a signature for the appearance
of anti-zeros has been given, and a conjecture made
relating this to the singular behaviour of certain 
elliptic curves. Further work needs to be made on
checking this conjecture with other examples, 
on proving that anti-zeros do exist, and on
finding indications for their existence in other
approaches, such as rational maps and spectral curves.

\newpage

\noindent{\bf Acknowledgements}

Many thanks to Conor Houghton for useful discussions.

\newpage
{\bf Figure Captions}\\

Fig 1. Components of the Higgs field for the
tetrahedral monopole.\\

Fig 2. The square of the Higgs field for the
cubic monopole (solid line) and two other configurations
(dashed lines) on the same geodesic.\\

Fig 3. As Fig 2. but for the component $\varphi_2$.\\

Fig 4. The square of the Higgs field for the
octahedral monopole.\\

\newpage
\begin{figure}[ht]
\begin{center}
{\Large \bf Fig. 1}
\end{center}
\vskip -3cm
\hskip 2cm {\epsfxsize=8cm \epsffile{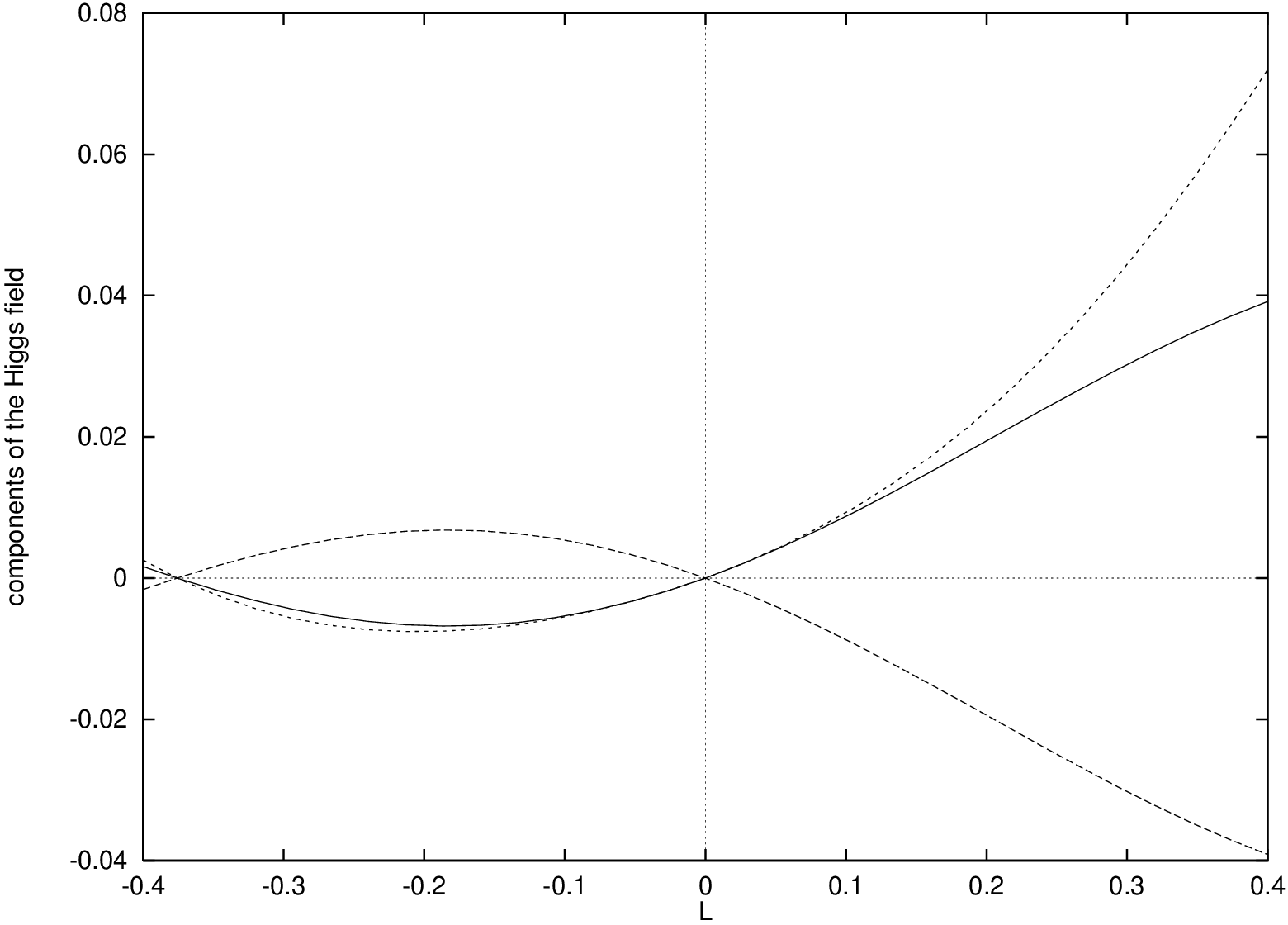}}
\begin{center}
{\Large \bf Fig. 2}
\end{center}
\vskip -3cm
\hskip 2cm {\epsfxsize=8cm \epsffile{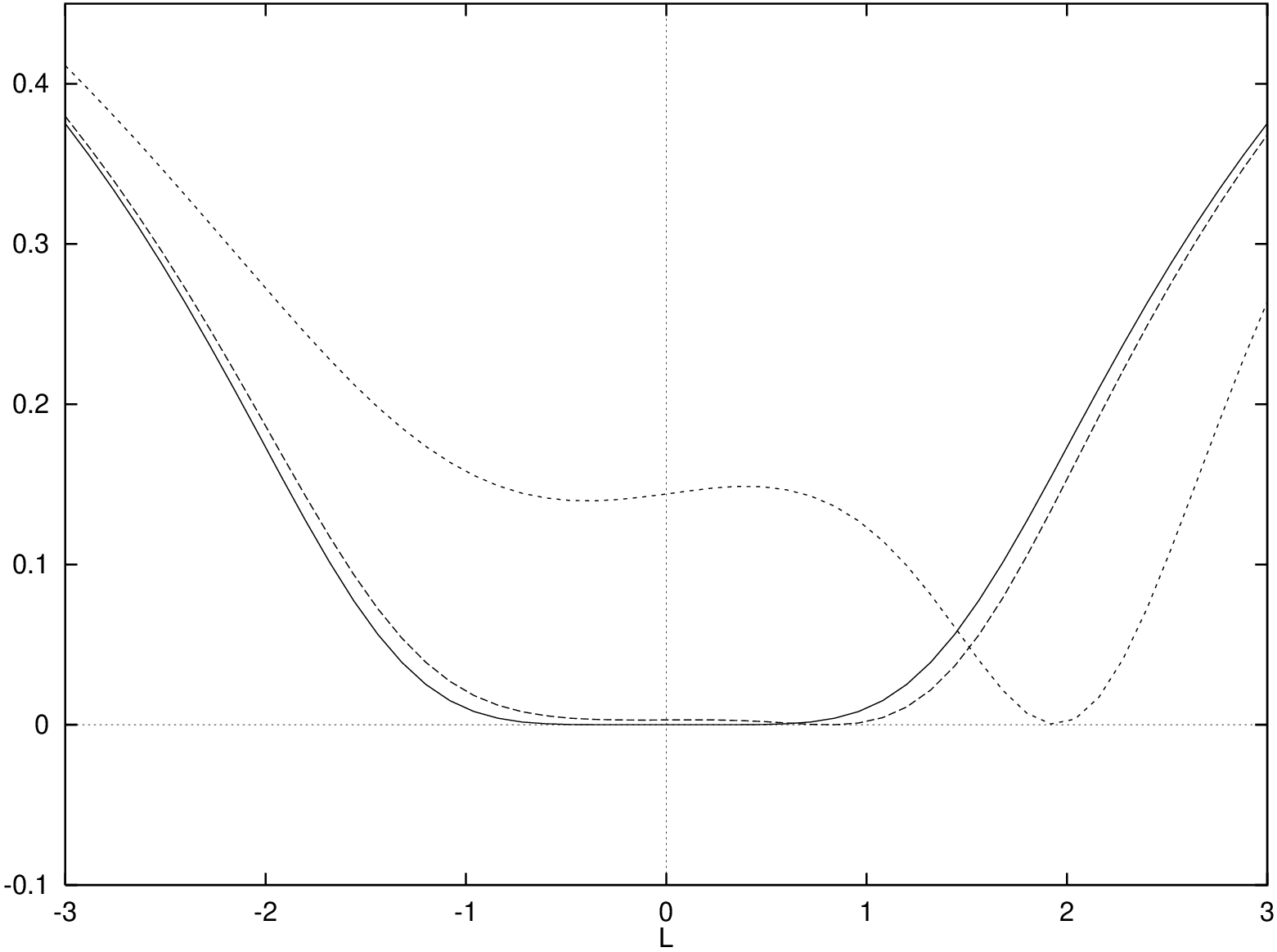}}
\end{figure}
\newpage
\begin{figure}[ht]
\begin{center}
{\Large \bf Fig. 3}
\end{center}
\vskip -3cm
\hskip 2cm {\epsfxsize=8cm \epsffile{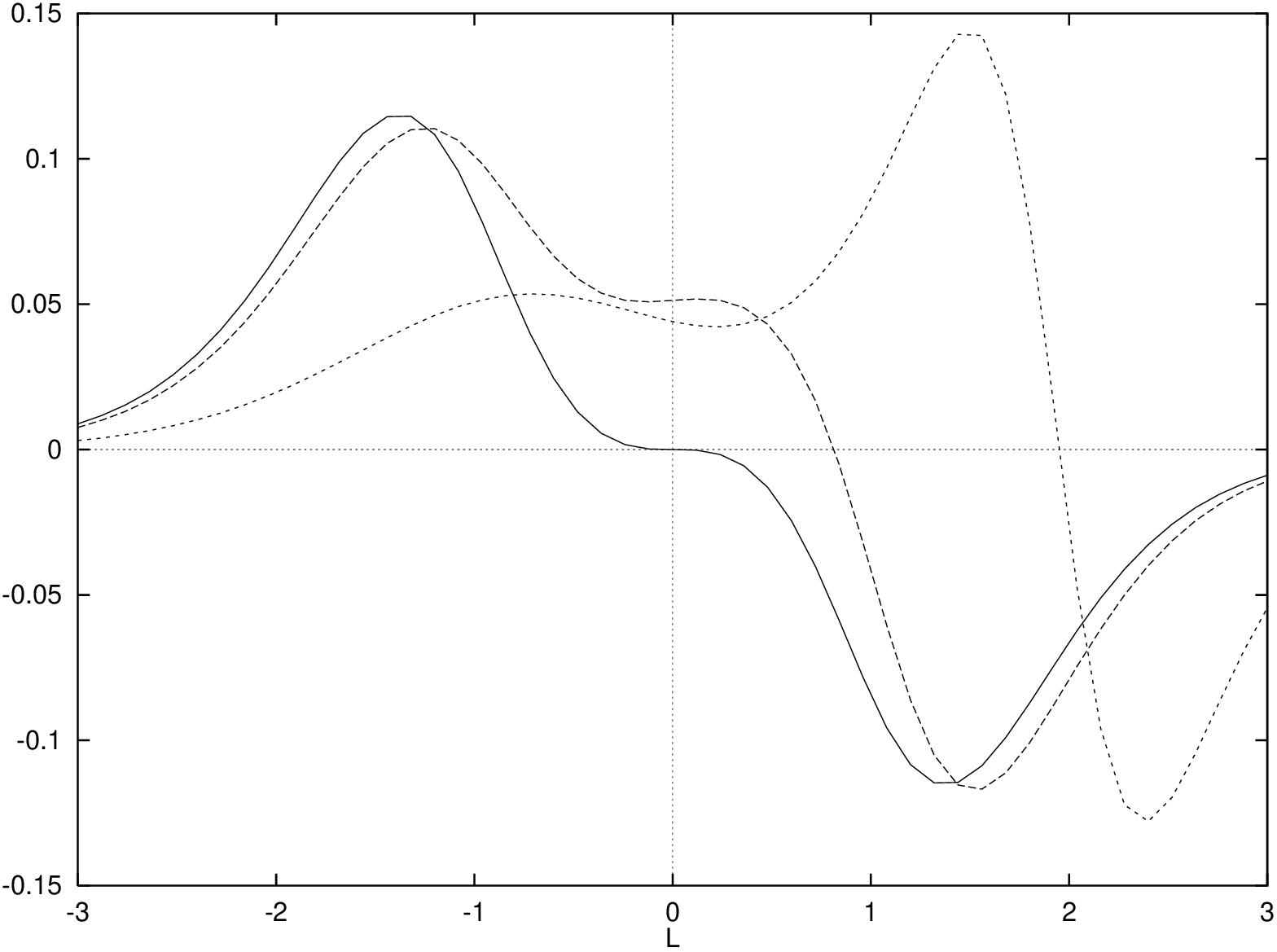}}
\begin{center}
{\Large \bf Fig. 4}
\end{center}
\vskip -3cm
\hskip 2cm {\epsfxsize=8cm \epsffile{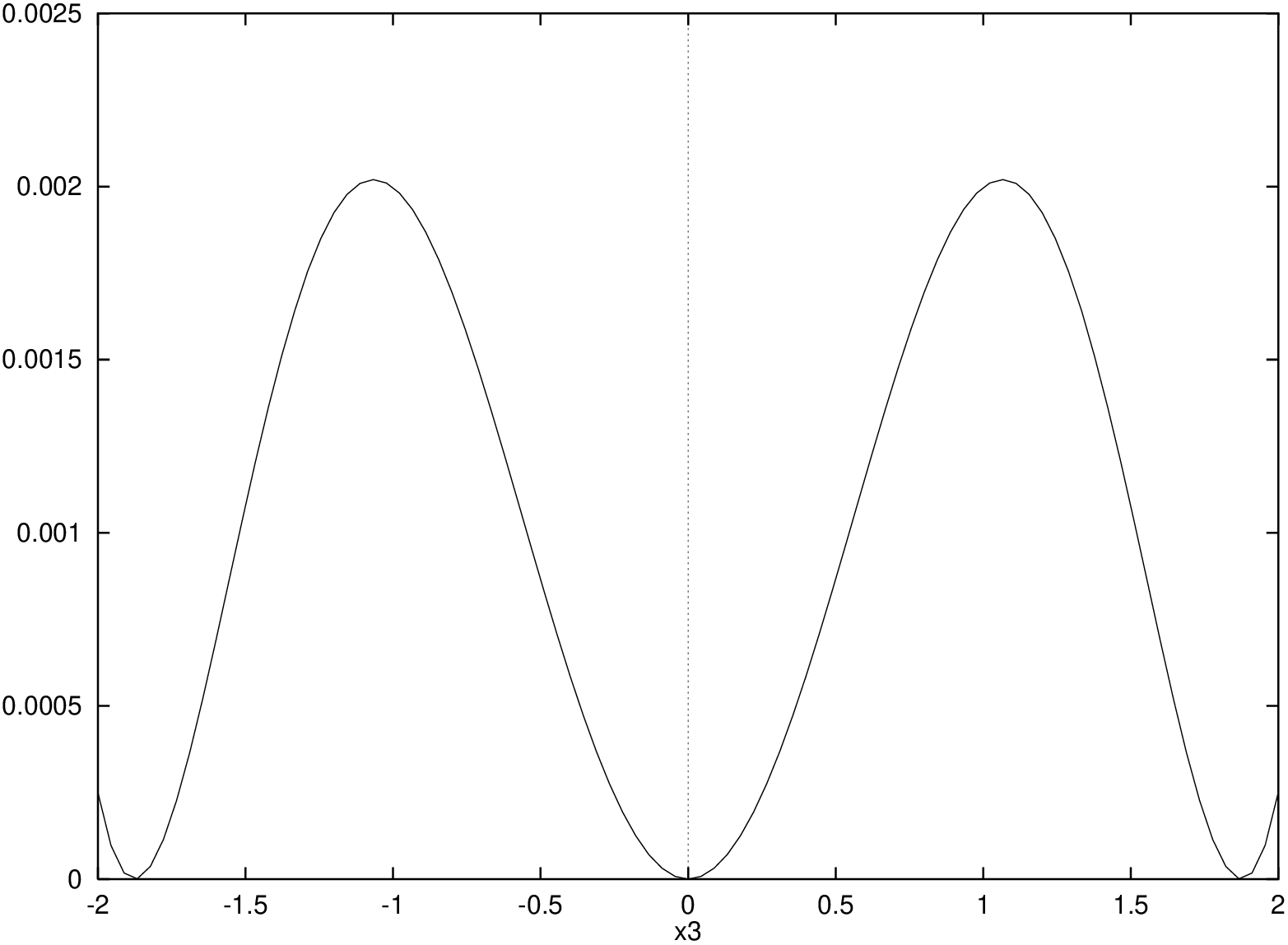}}
\end{figure}
\end{document}